\documentclass[aps,prl,twocolumn,superscriptaddress,amsfont,graphicx,nofootinbib,preprintnumbers]{revtex4}%
\usepackage{color,graphicx,epsfig}
\usepackage{ifpdf}
\usepackage{amsmath}
\usepackage{bm}
\usepackage{color}
\usepackage[english]{babel}
\usepackage{graphicx}%
\usepackage{amsfonts}%
\usepackage{amssymb}
\usepackage{braket}
\usepackage{hyperref}

\definecolor{nicered}{rgb}{0.7,0.1,0.1}
\definecolor{nicegreen}{rgb}{0.1,0.5,0.1}
\hypersetup{colorlinks,citecolor= nicegreen,linkcolor= nicered}

\newcommand{\beq}{\begin{equation}}
\newcommand{\eeq}{\end{equation}}
\newcommand{\bea}{\begin{eqnarray}}
\newcommand{\eea}{\end{eqnarray}}

\definecolor{Red}{rgb}{1.,0.,0.}

\def\mysection#1{{{\bf #1}.~}}
\begin{document}

\preprint{DO-TH 17/04}
\title{$R_K$ and $R_{K^{\ast}}$  beyond the Standard Model}
\author{Gudrun Hiller}
\email[Electronic address:]{ghiller@physik.uni-dortmund.de}
\author{Ivan Ni\v sand\v zi\'c}
\email[Electronic address:]{ivan.nisandzic@tu-dortmund.de}
\affiliation{Institut f\" ur Physik, Technische Universit\" at Dortmund, D-44221
Dortmund, Germany}

\begin{abstract}
Measurements of the ratio of  $B \to K^* \mu \mu $ to $B \to K^* e e $ branching fractions, $R_{K^*}$,  by the LHCb collaboration strengthen the hints
 from previous studies with pseudoscalar kaons, $R_K$,
for the breakdown of lepton universality, and therefore the Standard Model (SM),  to $\sim 3.5 \sigma$.
Complementarity between $R_K$ and $R_{K^*}$ allows to pin down the Dirac structure  of the new contributions to be
predominantly
SM-like chiral, with possible admixture of chirality-flipped contributions of up to ${\cal{O}}( \mbox{few} 10 \%)$.
Scalar and vector leptoquark representations $(S_3,V_1,V_3)$ plus  possible ($\tilde S_2,V_2$) admixture can  explain
$R_{K,K^*}$ via tree level exchange.
Flavor models naturally predict leptoquark masses not exceeding a few TeV, with 
couplings to third generation quarks at  $O(0.1)$, implying that this scenario can be directly tested  at the LHC.
\end{abstract}

\maketitle

\mysection{Introduction}
Gauge interactions of the leptons within the Standard Model (SM) exhibit exact universality. The only known source of  lepton non-universality (LNU) are the Yukawa couplings of the leptons to the Higgs. 
Tests of lepton universality are provided by  rare (semi)leptonic $|\Delta B|=|\Delta S|=1$ transitions, which are induced  in the SM at one loop  and additionally suppressed by the  Glashow-Iliopoulos-Maiani mechanism, therefore  allowing to probe  physics from  scales significantly higher than the weak scale.
Useful observables are the ratios of branching fractions of $B$ meson decays into strange hadrons $H$ and muon pairs over  electron pairs~\cite{Hiller:2003js}
\begin{align}
R_H=\frac{\mathcal{B}(B\to H \mu^+\mu^-)}{B(B\to H e^+e^-)} \, ,\quad \quad H=K,K^*,X_s, \ldots
\end{align}
in which (lepton universal)  hadronic effects cancel. The ratios are therefore predicted within the SM  to be very close to one and provide a clean
test of the SM~\cite{Hiller:2003js}. 

The LHCb collaboration measured $R_K$ in the dilepton invariant mass squared ($q^2$) bin $1 \,\text{GeV}^2 \leq q^2 \leq 6\,\text{GeV}^2$ using the $1\,\text{fb}^{-1}$ data set~\cite{Aaij:2014ora}
\begin{align}
R^{\text{LHCb}}_K&=0.745^{+0.090}_{-0.074}\pm 0.036 \, , \label{RKLHCb}
\end{align}
and, very recently, $R_{K^\ast}$
 within  $1.1 \,\text{GeV}^2 \leq q^2 \leq 6\,\text{GeV}^2$~\cite{LHCb2017}
\begin{align}
R^{\text{LHCb}}_{K^\ast}&=0.69^{+0.11}_{-0.07} \pm 0.05   \, , \label{RKstLHCb}
%0.685^{+0.113}_{-  0.069} \pm 0.047  \, , \label{RKstLHCb}
\end{align}
with  deviation from $R=1$ by $2.6\,\sigma$ each. (Here and in the following we add statistical and systematic uncertainties in quadrature.)
Corrections to $R=1+O(m_\mu^2/m_B^2)$~\cite{Hiller:2003js} arise from 
electromagnetic interactions~\cite{Huber:2005ig, Bobeth:2007dw,Huber:2007vv,Huber:2015sra}.
This affects the SM prediction at low $q^2$ at percent level~\cite{Bordone:2016gaq}, not qualitatively altering the fact that the data, (\ref{RKLHCb}) and (\ref{RKstLHCb}) constitute a
challenge to universality, and the SM.

Moreover, the importance of the measurement of $R_{K^*}$ in addition to $R_K$ is in its diagnosing power regarding different  beyond the SM (BSM) contributions~\cite{Hiller:2014ula}.
Left-handed  and right-handed  $b \to s$ currents
enter  $B \to K \ell \ell$ and $B \to K^* \ell \ell$ in almost  orthogonal combinations in both regions of $q^2$  sensitive to LNU. Comparison of
$R_K$ with $R_{K^*}$,  for instance through a double ratio  $X_{K^\ast}=R_{K^\ast}/R_K$~\cite{Hiller:2014ula}, probes directly right-handed LNU currents.  
The aim of this paper is to exploit this  model-independently and pursue interpretations within leptoquark  extensions of the SM.

\mysection{Model-independent interpretation\label{sec:MIA}}
We employ the usual effective Hamiltonian for  $b\to s\ell\ell$, $\ell=e,\mu, \tau$ transitions  
\begin{equation}
\mathcal{H}_{\text{eff}}=-\frac{4G_F \lambda_t}{\sqrt{2}}\frac{\alpha}{4\pi}\sum_{i} C^{\ell}_i\mathcal{O}^{\ell}_i+\text{h.c.},
\end{equation}
where $C^{\ell}_i, \mathcal{O}^{\ell}_i$ denote lepton-specific Wilson coefficients and dimension-six  operators, respectively, 
renormalized at the  scale $\mu\sim m_b$. Furthermore, $G_F$, $\alpha$, and $\lambda_t=V_{tb}V^*_{ts}$  stand for Fermi's constant, the finestructure  constant and the product of relevant Cabibbo-Kobayashi-Maskawa (CKM) matrix elements, respectively.
The semileptonic operators read
\begin{equation}
\begin{split}
\mathcal{O}^{\ell}_9&=(\bar{s}\gamma^\mu P_L b)(\bar{\ell}\gamma_\mu \ell),\quad \mathcal{O}'^{\ell}_9=(\bar{s}\gamma^\mu P_R b)(\bar{\ell}\gamma_\mu \ell),\\
 \mathcal{O}^{\ell}_{10}&=(\bar{s}\gamma^\mu P_L b)(\bar{\ell}\gamma_\mu\gamma_5 \ell),\quad
\mathcal{O}'^{\ell}_{10}=(\bar{s}\gamma^\mu P_R b)(\bar{\ell}\gamma_\mu\gamma_5 \ell),
\end{split}
\end{equation}
with chiral projectors $P_{L,R}=1/2(1\mp \gamma_5)$.
The operators with chiral  lepton currents, 
\begin{equation}
\mathcal{O}^\ell_{AB}=(\bar{s}\gamma^\mu P_A b)(\bar{\ell}\gamma_\mu P_B \ell) \, , \quad \quad A,B=L,R \, ,
\end{equation}
are related to the ${\cal{O}}_{9,10}^{(\prime) \ell}$ as
\begin{equation}
\begin{split}
C^{\ell}_{9}&=\frac{1}{2}(C^{\ell}_{LL}+C^{\ell}_{LR}),\quad C^{\ell}_{10}=\frac{1}{2}(C^{\ell}_{LR}-C^{\ell}_{LL}) \, , 
\\
C'^{\ell}_{9}&=\frac{1}{2}(C^{\ell}_{RL}+C^{\ell}_{RR}),\quad C'^{\ell}_{10}=\frac{1}{2}(C^{\ell}_{RR}-C^{\ell}_{RL}) \, . 
\end{split}
\end{equation}

Within the SM the (lepton universal) Wilson coefficients  are $C^{\text{SM}}_9=4.07$, $C^{\text{SM}}_{10}\simeq -4.31$ \cite{global fits}, thus $C^{\rm SM}_{LL}=C^{\text{SM}}_9-C^{\text{SM}}_{10}\simeq 8.4$, while   scalar or tensor Wilson coefficients are negligible. 
We define $C^{\ell}_{LL}=C^{\rm SM}_{LL}+C^{\text{NP}\ell}_{LL}$~\cite{Hiller:2014ula} and drop the label "NP" (new physics)  for Wilson coefficients negligible within the SM. 

In the $B\to K^{(\ast)}\ell\ell$  branching fractions  contributions from photon exchange enter, notably from
charm loops and dipole operators. These contributions are numerically small at high and low $q^2$,  sufficiently  away  from the photon pole, and
lepton universal. Within current accuracy of $R_{K,K^*}$  these contributions can be safely neglected. In this limit~\cite{Hiller:2014ula}
\begin{equation}
\begin{split}
R_K&=1+\Delta_++\Sigma_+,\quad\\ R_{K^{\ast}}&=1+\Delta_++\Sigma_++p(\Sigma_--\Sigma_++\Delta_--\Delta_+),\label{RK,RKst}
\end{split}
\end{equation}
where 
\begin{equation}
\begin{split}
\Delta_{\pm}&=2 \Re\bigg(\frac{C^{\rm NP\,\mu}_{LL}\pm C^\mu_{RL}}{C^{ \rm SM}_{LL}}-(\mu \rightarrow e)\bigg) \, , \\
\Sigma_{\pm}&=\frac{\vert C^{ \rm NP\,\mu}_{LL}\pm C^{\mu}_{RL}\vert^2+\vert C^{\mu}_{LR}\pm C^{\mu}_{RR}\vert^2}{\vert C^{\rm SM}_{LL}\vert^2}-(\mu \rightarrow e).
\end{split}
\label{DeltaSigma}
\end{equation}
Since BSM contributions in  $|\Delta B|=|\Delta S|=1$ transitions are smaller than the SM ones \cite{global fits}  the dominant BSM effect is captured by the  linear (interference) terms $\Delta_\pm$.

The coefficient $p$  in Eq.~\eqref{RK,RKst} denotes the fraction of transverse parallel and longitudinal  contributions to the $B\to K^\ast \ell\ell$ branching ratio~\cite{Hiller:2014ula}.
Due to helicity arguments $p \sim 1$ both at low recoil (high $q^2$) and at low $q^2$. Consequently, ${\cal{B}}(B\to K^\ast \ell\ell)$ is dominated by
contributions proportional to $ |C -C^\prime|^2$. Since ${\cal{B}}(B\to K \ell\ell)\propto |C+C^\prime|^2$ due to parity invariance of the strong interaction  both modes are complementary and deviations of  $R_K$ from $R_{K^*}$ probe primed operators~\cite{Hiller:2014ula}.

Using  (\ref{RKLHCb}),(\ref{RKstLHCb}) one obtains
\begin{align}
X_{K^\ast} =R_{K^\ast}/R_K& =0.94 \pm 0.18      \, , \label{XKst} \\
R_{K^\ast}+R_K-2&=-0.54 \pm 0.14      \, , \label{sum}
\end{align}
which gives,  at $1 \sigma$ 
\begin{align} \label{eq:XlimitL}
 {\rm Re}[ C_9^{{\rm NP} \mu}  -
C_{10}^{{\rm NP}  \mu}  -(\mu \to e)  ] &  \sim -1.1 \pm 0.3  \, , \\
  {\rm Re}[ C_9^{\prime  \mu}  - C_{10}^{\prime \mu}  -(\mu \to e)  ] &  \sim  0.1 \pm 0.4  \, .
   \label{eq:XlimitR}
\end{align}
As anticipated,  $|C^{\rm NP}| \ll |C^{\rm SM}|$. Therefore, the linear approximation, that is, neglecting the $\Sigma_\pm$-terms, is meaningful within the current experimental precision.
Dropping quadratic terms greatly simplifies the interpretation of the data: Only BSM in $O_{LL}^\ell$ or $O_{RL}^\ell$ is able to  explain $R_{K,K^*}$.

\begin{figure}[t]
\centering{
\includegraphics[height=0.27\textwidth]{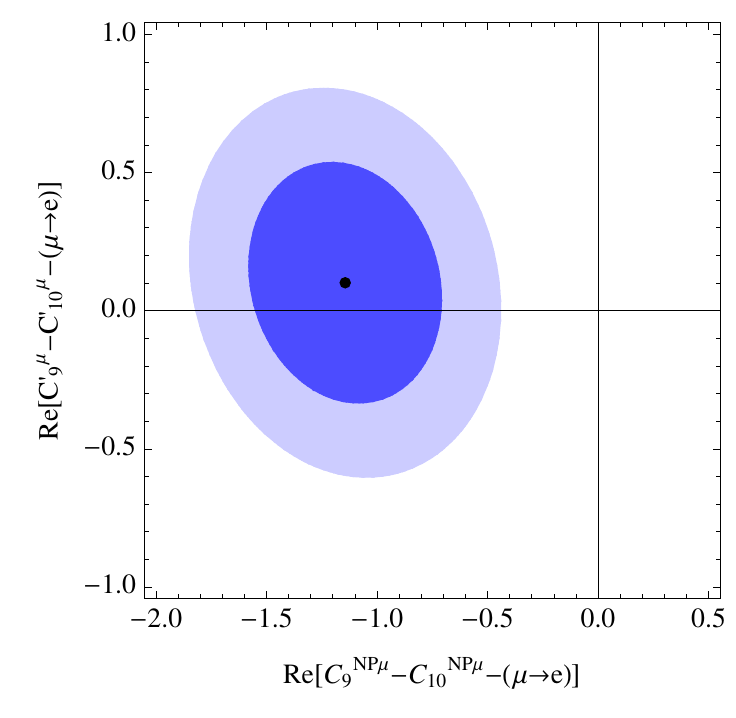}
}
\caption{Fit of left- and light-handed BSM coefficients  in $|\Delta B|=|\Delta S|=1$ transitions to  $R_K$ and $R_{K^*}$ data (\ref{RKLHCb}), (\ref{RKstLHCb}).
Darker and lighter shaded regions correspond to 68  and 95 \% CL intervals, respectively.}
\label{fig:chi2}
\end{figure}

In Fig.~\ref{fig:chi2} a $\chi^2$ fit for the left- and right-handed  Wilson coefficients is shown. The discrepancy with the SM is
about $\sim 3.5 \sigma$, where we allowed for a few percent deviations from $R=1$ \cite{Bordone:2016gaq}.
Interestingly, solutions with $C_9^{ {\rm NP} \mu} \sim -1$ are also favored by a global fit~\cite{global fits} to  $b\to s\mu\mu$ observables.
Taking this into account suggests an explanation of $R_{K,K^*}$ anomalies with BSM predominantly residing in the muons.

\mysection{Leptoquark explanations}
We consider  leptoquark extensions of the SM  with tree level couplings to down-type quarks and leptons.
 Representations under $SU(3)_C \times SU(2)_L \times U(1)_Y$ with relevant Wilson coefficients are given in Table~\ref{table:LQs} for scalar $S_i$\footnote{In the literature the scalar leptoquarks $S_2$ and $\tilde S_2$ are also denoted  by $R_2$ and $\tilde R_2$, respectively.} and 
 in Table~\ref{tab:Vector-LQs} for vector leptoquarks $V_i$, respectively. The index $i=1,2,3$ refers to the dimension of the $SU(2)_L$ multiplet,  see {\it e.g.}~\cite{Kosnik:2012dj,Hiller:2016kry,Dorsner:2016wpm} for  overviews.

The scalar leptoquarks $S_2$ and $\tilde{S}_1$  generate only $C_{LR}$ and $C_{RR}$, respectively, which do not interfere with the SM contribution, see Eq.~\eqref{DeltaSigma}, and lead to $R_K, R_{K^\ast}$ near 1. We therefore discard these two possibilities as explanations of the $R_{K,K^\ast}$ anomalies. 
\begin{table}[t]
\centering
\renewcommand{\arraystretch}{1.6}
\title{Details of the scalar leptoquark scenarios}
 \begin{tabular}{l c c c c}\hline \hline
 & representation  & $C_{AB}$ &  Relation & $R_{K^{(\ast)}}$ \\\hline
$\tilde{S}_2$& $(3,2,1/6)$ &$C_{RL}$  & $C'_9=-C'_{10}$& $R_K<1,\,R_{K^\ast}>1$ \\                                
 \hline
$S_3$ & $(\bar{3},3,1/3)$&$C^{\text{NP}}_{LL}$& $C_9=-C_{10}$ & $R_K \simeq R_{K^\ast}<1.$ \\  
\hline
$S_2$& $(3,2,7/6)$&$C_{LR}$&$C_9=C_{10}$&$R_K\simeq R_{K^\ast}\simeq 1$ \\
\hline
$\tilde{S}_1$& $(\bar{3},1,4/3)$ &$C_{RR}$&  ${C}'_9={C}'_{10}$ & $R_K\simeq R_{K^\ast}\simeq 1$\\
\hline\hline
 \end{tabular}\caption{Scalar leptoquarks and  relations between  Wilson coefficients, assuming a single leptoquark at the time. The last column shows implications for $R_{K^\ast}$ assuming  $R_K<1$, as suggested by data \eqref{RKLHCb}. \label{table:LQs}}
 \end{table} 

 \begin{table}[t]
\centering
\renewcommand{\arraystretch}{1.6}
\title{Details of the vector leptoquark scenarios for $b\to s\ell\ell$}
 \begin{tabular}{l c c c c}\hline \hline
& representation  & $C_{AB}$&  Relation & $R_{K^{(\ast)}}$ \\\hline
$V_1$& $(3,1,2/3)$ &$C^{\text{NP}}_{LL}$  & $C_9=-C_{10}$& $R_K \simeq R_{K^\ast}<1$ \\    
      &                   &          $C_{RR}$             & $C'_9=+C'_{10}$   &   $R_K\simeq R_{K^\ast}\simeq 1$ \\
 \hline
$V_2$ & $(3,2,-5/6)$&$C_{RL}$& $C'_9=-C'_{10}$ & $R_K<1,\,R_{K^\ast}>1$ \\  
&                   &          $C_{LR}$             & $C_9=+C_{10}$   &   $R_K\simeq R_{K^\ast}\simeq 1$ \\
\hline
$V_3$& $(3,3,-2/3)$&$C^{\text{NP}}_{LL}$&$C_9=-C_{10}$&$R_K \simeq R_{K^\ast}<1$ \\
\hline\hline
 \end{tabular}\caption{Vector leptoquarks and implications for $R_{K^\ast}$ assuming  $R_K<1$, as  suggested by data \eqref{RKLHCb},  see Table \ref{table:LQs}.}
 \label{tab:Vector-LQs}
 \end{table} 

In view of the experimental constraints  shown in  Fig.~\ref{fig:chi2} we focus on leptoquarks that can give a sizable 
$C^{\text{NP} \ell}_{LL}=2 C_9^{\text{NP}\ell}=-2C_{10}^{\text{NP}\ell}$.
This singles out the scalar triplet $S_3$,
 the vector singlet $V_1$ and the vector triplet $V_3$. This scalar and the vectors have been  considered as a possible explanation of $R_K$ \eqref{RKLHCb} in ~\cite{Hiller:2014yaa, Gripaios:2014tna, Hiller:2016kry,Barbieri:2015yvd,Varzielas:2015iva} and   in~\cite{Barbieri:2015yvd,Fajfer:2015ycq,Becirevic:2016oho,Alonso:2015sja, Calibbi:2015kma, Hiller:2016kry}, respectively.
Subdominant contributions from right-handed currents can be provided   by additional leptoquarks $\tilde S_2$ or $V_2$, which induce $C_{RL}^\ell=2 C_9^{\prime \ell}=-2C_{10}^{\prime \ell}$.
In these models \cite{Dorsner:2016wpm,Hiller:2016kry}
\begin{align}
C^{\text{NP}\ell}_{LL} = &  \frac{ k_{LQ} \pi \sqrt{2}}{G_F \lambda_t \alpha} \frac{ Y  Y^*}{M^2} \, ,  k_{LQ}=1,-1,- 1\,  \mbox{for}  \,   S_3,V_1,V_3  ,\\
C_{RL}^\ell= &   \frac{  k_{LQ}\pi \sqrt{2}}{G_F \lambda_t \alpha} \frac{ Y Y^* }{M^2} \, ,  k_{LQ} =-1/2,+1 \,  \mbox{for} \, \,  \tilde S_2,V_2. 
\end{align}
Here, $M$  ($Y$) denotes the leptoquark mass (coupling).

\begin{figure}[t]
\centering{
\includegraphics[height=0.27\textwidth]{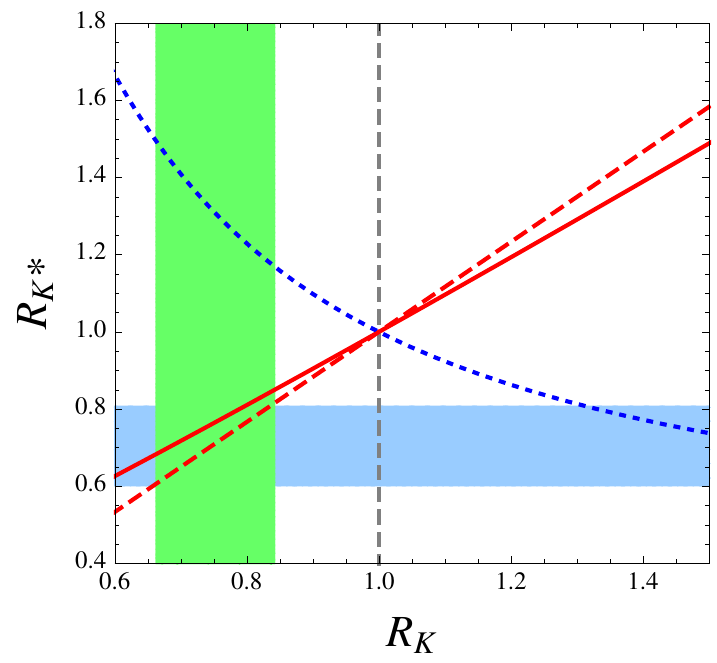}
}
\caption{$R_K$ versus  $R_{K^*}$ in BSM scenarios.
Solid red curve: $C^{\rm NP}_{LL}$ ($C^{\rm NP}_9=-C^{\rm NP}_{10}$) corresponding to $S_3,V_1$ or $V_3$, blue dotted curve: $C_{RL}$ ($\tilde S_2$ or $V_2$), gray dashed curve: $C_{RL}=-C^{\rm NP}_{LL}$ (no single leptoquark), and  red dashed curve: $C^{\rm NP}_{LL}$ and $C_{RL}=-1/10\, C^{\rm NP}_{LL}$ (for instance, $S_3$ plus 
$10\%$ admixture of $\tilde S _2$). The colored bands correspond to the LHCb measurements of $R_K$  (\ref{RKLHCb})  and $R_{K^\ast}$  (\ref{RKstLHCb}).}
\label{fig:correlation}
\end{figure}

Model-independent and leptoquark specific predictions for $R_K$ versus $R_{K^*}$ 
are shown in Fig.~\ref{fig:correlation}.  The green and blue  band  denote  the  $1 \sigma$ band of $R_K$ (\ref{RKLHCb}) and $R_{K^*}$ (\ref{RKstLHCb}),
   respectively.
Also shown are BSM scenarios which can (red solid and dashed lines) or cannot (blue dotted  and gray dashed lines) simultaneously explain the data.
Concretely, leptoquark $\tilde S_2$, corresponding to the blue dotted curve, and which has been considered in the context of $R_K$ \cite{Hiller:2014yaa,Sahoo:2015wya,Becirevic:2015asa,Cox:2016epl}, is disfavoured as the sole source of LNU  by  the measurement of $R_{K^*}$. The numerics  are based on the full expressions for the decay rates,
for $\ell=\mu$. Recall, however, to  linear approximation only non-universality matters.

We find for the dominant, SM-like chiral contribution $S_3$
\begin{equation} \label{eq:S3}
\frac{ Y_{b\mu} Y^\ast_{s\mu}-  Y_{be} Y^\ast_{se} }{M^2} \simeq \frac{ 1.1}{  (35\,\text{TeV})^2} \, ,  \quad \quad (S_3)
\end{equation}
and similarly for $V_1$ or $V_3$.
To accommodate an admixure of right-handed currents we need contributions from another leptoquark, such as $\tilde S_2$
\begin{equation}
\frac{ Y_{b\mu} Y^\ast_{s\mu}  -  Y_{be} Y^\ast_{se}}{M^2} \simeq \frac{ -0.1}{  (24\,\text{TeV})^2}. \quad \quad (\tilde S_2)
\end{equation}
Understanding the mass range is linked to flavor.
The leptoquark coupling matrix $Y$ is a $3 \times 3 $ matrix in generation space, with rows corresponding to quark flavor and columns corresponding to lepton flavor.
The presence of both kinds of fermions in one vertex is  beneficial; it allows to probe flavor in new ways  beyond SM fermion masses and mixings.
Viable models are those employing a Froggatt-Nielsen $U(1)_{FN}$ to generate mass hierarchies for quarks and charged leptons together with a discrete, non-abelian group such as $A_4$, which
allows to accommodate neutrino properties \cite{Altarelli:2010gt,Varzielas:2010mp}.
Applied  to leptoquark models this allows to select lepton species --  for instance having  only couplings to one lepton species, muons, or electrons \cite{Varzielas:2015iva}.
Corrections to lepton isolation arise from rotations to the mass basis and at higher order in the spurion expansion, and induce lepton flavor violation 
\cite{Glashow:2014iga,Becirevic:2016zri,Varzielas:2015iva, Hiller:2016kry,Duraisamy:2016gsd,Crivellin:2017zlb} such as $B\to K\mu \tau$, which can be probed with $B$-physics experiments but also $\mu-e$-converison, rare $K$ and $\ell \to \ell^\prime$ decays.
Together with $B \to K^{(*)} \nu \bar \nu$ modes the  latter  constitute the leading constraints on flavor models and LNU anomalies, and improved experimental study is promising.

A generic prediction for $S_3,V_1,V_3$ -- all of them couple quark doublets to lepton doublets--  is obtained from 
simple flavor patterns such as $\ell$-isolation, $\ell=e ,\mu$,   \cite{Varzielas:2015iva,Hiller:2016kry}
\begin{equation}  \label{eq:flavor}
   Y_{q_3 \ell} \sim c_l\, , \quad   Y_{q_2 \ell} \sim c_l \lambda^2 \, , \quad  q_3=b,t,\,  ~ q_2=s,c \, , 
\end{equation}
where $c_l \sim  \lambda \sim 0.2$. Note that the FN-mechanism is only able  to explain parametric suppressions in  specific powers of the parameter $\lambda$
up to numbers of order one. Irrespective of the concrete flavor symmetry,
each coupling $Y$ to lepton doublets brings in a  non-abelian spurion insertion  suppression, the factor $c_l$, which is  unavoidable as lepon doublets are necessarily  charged under the non-abelian group to 
obtain a viable PMNS-matrix.
The suppression of the additional couplings to  right-handed leptons in $V_{1,2}$ can be achieved using  flavor symmetries~\cite{Alonso:2015sja, Hiller:2016kry}.

Putting lepton and neutrino properties aside,
a minimal prediction  is $Y_{s \ell}/Y_{b \ell} \sim m_s/m_b $, hence
$Y_{b \ell} Y_{s \ell}^*  \sim \lambda^2  \simeq \mbox{ few} \times  0.01$. Eq.~(\ref{eq:S3})   implies $M \sim 5-10$ TeV, 
accessible  at the LHC at least partly  with single  production.

\begin{figure}[t]
\centering{
\includegraphics[height=0.27\textwidth]{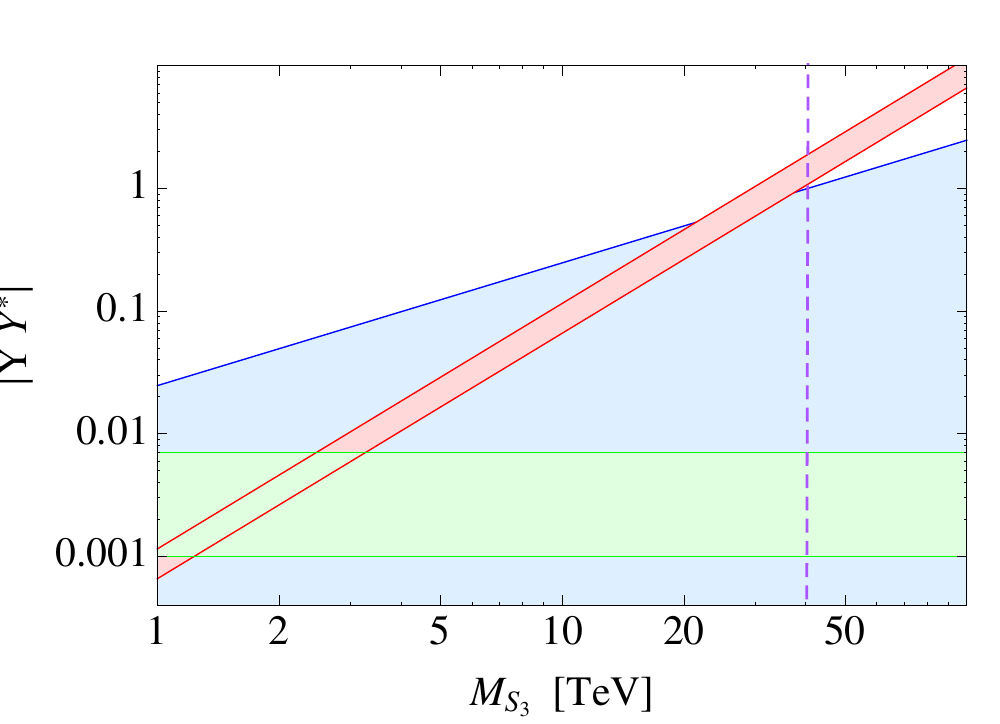}
}
\caption{Allowed values of $|Y  Y^*|, M_{S_3}$  by  $\Delta m_{B_s}$ (blue area) and  $R_{K^{(\ast)}}$ (red band) \eqref{eq:XlimitL}. The  green band corresponds to  flavor model predictions (\ref{eq:flavor}). The dashed  blue line corresponds to the upper limit on the mass of the $S_3$ leptoquark  (\ref{eq:masslimit}).}
\label{fig:mass}
\end{figure}

Eq.~(\ref{eq:flavor}) points to lower values of leptoquark masses, see Fig.~\ref{fig:mass}.
Also shown are constraints from $B_s -\bar B_s$ mixing, induced at one loop  through box diagrams and  which constrains the  {\it square} of  $Y Y^*$ over $M^2$
\cite{Hiller:2014yaa}.
A data-driven upper  limit, irrespective of flavor, is obtained as
\begin{align} \label{eq:masslimit}
M \lesssim 40 \, \mbox{TeV} \, , 45 \, \mbox{TeV} \, , 20 \, \mbox{TeV}\quad \quad  \mbox{for}~ S_3, V_1, V_3 \, . 
\end{align}
We assume vector leptoquarks to be gauge-like and employ the usual 
Hamiltonian 
\begin{equation}
\mathcal{H}^{\Delta B=2}_{\text{eff}}=(C_1^{SM}+C_1^{LQ})(\bar{b}\gamma_\mu(1-\gamma_5)s)(\bar{b}\gamma_\mu(1-\gamma_5)s)+\text{h.c.}\label{DeltaB-Hamiltonian}
\end{equation}
where
\begin{equation}
C_1^{LQ}=\frac{ p_{LQ} (Y Y^*)^2}{128 \pi^2 M^2} \,  , ~ p_{LQ}=5, 4, 20  \, \,  \mbox{for} \, \, S_3,V_1,V_3  \, ,
\end{equation}
see, {\it e.g}~\cite{Davidson:1993qk}. In general,  $(Y Y^*)^2\to \sum_{\ell_i,\ell_j}(Y_{b\ell_i}Y^\ast_{s\ell_i})(Y_{b\ell_j}Y^\ast_{s\ell_j})$.
It follows that 
\begin{equation}
\Delta m^{\text{LQ}}_{B_s}/\Delta m_{B_s}^{SM}=\frac{p_{LQ}(Y Y^*)^2}{8 M^2 G_F^2 m_W^2 \lambda_t^2 S_0(x_t)},
\end{equation}
where $S_0$ is an Inami-Lim function, $x_t=m_t^2/m_W^2$. We use $\Delta m_{B_s}^{exp}/\Delta m_{B_s}^{SM}=1.02\pm 0.10$~\cite{Olive:2016xmw,Becirevic:2016zri}.

Direct limits for scalar leptoquarks decaying  100 \% into a muon (electron) and a jet are are $M > 1050$ GeV \cite{Aaboud:2016qeg} ($M>1755$ GeV \cite{Khachatryan:2015qda}).
For vector leptoquarks, the limits  are model-dependent and read 
 $M > 1200 -1720$ GeV ($M > 1150 -1660$ GeV) for 100 \% decays to muon (electron) plus jet \cite{Khachatryan:2015vaa}.
 The bounds weaken if decays into neutrinos are taken into account.
 
\mysection{UV considerations}
The main challenge for embedding  light scalar leptoquarks into (complete) short-distance models is  proton decay. From Table~\ref{table:LQs} only $S_2, \tilde{S}_2$ do not couple to quark bilinears $(\bar{q}q)$ and, thus, do not induce proton decay at tree-level. 
In addition, dangerous couplings to the Higgs doublet should be suppressed~\cite{Arnold:2012sd,Dorsner:2016wpm}.

SM gauge-invariance allows $S_3$ to couple  to quark billinears
\begin{equation}
\mathcal{L}_{QQ}=Y_\kappa\,\bar{Q}^{C\,\alpha}_L(i\,\sigma^2)^{\alpha\beta}(S_3^\dagger)^{\beta\gamma}Q_L^\gamma+\text{h.c.}, \label{diquarks}
\end{equation}
with isospin indices $\alpha, \beta,\gamma$.
The Yukawa coupling $Y_\kappa$ is  anti-symmetric in flavor space~\cite{Vecchi:2011ab, Dorsner:2012nq}, thus $S_3$ does not introduce  proton decay at tree level, however, couplings to $u t(c)$ can induce the process via higher orders diagrams~\cite{Dorsner:2012nq}.

Within flavor models, the dangerous terms in (\ref{diquarks}) receive suppressions. For  $U(1)_{FN} \times A_4 \times Z_3$,
and assuming that the quarks transform trivially under $A_4$, we find that this requires at least 2 spurion fields $\xi$ and $\xi^\prime$, see~\cite{Varzielas:2015iva} for details. Including the FN-suppression for the up-quark this amounts to
$\lambda^4 \kappa \kappa^\prime \lesssim  10^{-4}$. Viable patterns for $R_K$ are obtained with second generation quarks in non-trivial representations of $A_4$.
This way, however, the $ut$ coupling cannot be suppressed further.  

If the evidence for LNU in $C^{\text{NP}}_{LL}$ strengthens, it would be important to understand the origin of the leptoquark $S_3$ which provides an explicit high-energy realization. One possibility was suggested in Ref.~\cite{Gripaios:2014tna}. The $S_3$ appears, along with the Higgs doublet, as a pseudo-Goldstone boson of the strong dynamics around TeV scale, while the proton decay is avoided with a discrete symmetry.

An alternative possibility is to trace the origin of $S_3$ to a Grand Unified Theory (GUT).  The $S_3$ is contained in the $\overline{126}_H$ scalar multiplet of $SO(10)$~\cite{Slansky:1981yr, Dorsner:2016wpm}. The dangerous couplings to quark billinears are forbidden by  $SO(10)$-invariance - the corresponding Yukawa coupling to fermion multiplets is $y_{ij}\,16_i\,16_j\,\overline{126}_H$ which embeds only the couplings to leptons and quarks, but  not  to quark billinears. The $16_i$ denotes the spinor representation of $SO(10)$ that unifies all SM fermions of a single generation and a right-handed neutrino, and $i=1,2,3$ is a flavor index. The $S_3$ might play a role in correcting the phenomenologically unsuccessful prediction of the relation between the mass matrices of down-type quarks and charged leptons in the minimal $SU(5)$~\cite{Georgi:1979df, Georgi:1974sy}.

Vector leptoquarks appear as super-heavy gauge bosons in a GUT, with masses near the unification scale. For example, the state with quantum numbers of $V_1$ is a gauge boson in models of quark-lepton unification, {\it e.g.}~the original Pati-Salam model or variants thereof, see~\cite{Perez:2013osa}. $V_1, V_3$ do not couple to quark billinears and are safe with regards to proton decay. If $V_1$ is a gauge boson, the corresponding left- and right-handed couplings are unitary. It is then more difficult to suppress the unwanted (right-handed) couplings and simultaneously avoid the constraints from the first generation  fermions. The embedding of $V_3$ into a UV complete model is  challenging~\cite{Biggio:2016wyy}.

The low scale non-gauge spin-1 leptoquarks might be obtained as composite states from  strongly coupled dynamics, in which case they are accompanied by  other composite states. 

\mysection{Conclusions}
The recent measurement  of  $R_{K^*}$   (\ref{RKstLHCb})  by the LHCb collaboration challenges lepton universality, an inbuilt feature of the SM and many of its extensions, further: combined with
$R_K$ (\ref{RKLHCb})  the  discrepancy with the SM is $\sim 3.5 \sigma$.
The LNU contributions to $|\Delta B|=|\Delta S|=1$ FCNCs are predominantly SM-like chiral, with possible admixture of right-handed contributions
up to the order of few $10 \%$, see  Fig.~\ref{fig:chi2}. Since $R_K$ and $R_{K^*}$ suffice to determine the chiral structure, measurements of further  LNU ratios $R_H$ into different final states
 and angular distributions \cite{Hiller:2014ula,Wehle:2016yoi} provide consistency checks.
Using  $R_{K,K^*}$ data  we predict the ratio of inclusive $B \to X_s \ell \ell$ branching fractions
\begin{align}
R_{X_s} \sim 0.73 \pm 0.07 \, , 
\end{align}
consistent with earlier findings by Belle $R_{X_s} =0.42 \pm 0.25$    \cite{iijimaLP09}  and BaBar $R_{X_s} =0.58 \pm 0.19$   \cite{Lees:2013nxa} \footnote{The BaBar collaboration finds an excess of electrons relative to the SM prediction, especially in the lowest $q^2$-bin. }.
%%%Predictions for $R_\phi, R_{K_1}, R_{K_0(1430)}$ are given in \cite{Hiller:2014ula}.

Leptoquarks naturally induce LNU in semileptonic decays at tree level. The scalar  $S_3$ and the vector $V_{1,3}$ representations can  account for the
dominant, SM-like chiral contribution (\ref{eq:XlimitL}).
Their masses are limited to not exceed the multi-10 TeV range in order to comply with data, see (\ref{eq:masslimit}) for details.
Leptoquark explanations  of $R_{K,K^*}$ within flavor models, which simultaneously address the masses and mixings of SM fermions require leptoquark masses in the few TeV-region, which
can be explored at the LHC, see Fig.~\ref{fig:mass}. The dominant decay modes of  the triplets $S_3$ and $V_3$ are $b \mu, t \mu, b \nu$ and $t \nu$, whereas 
the $SU(2)_L$-singlet  $V_1$ decays predominantly  to $b \mu,  t \nu$. The respective Yukawa couplings are at the level $O(0.1)$.
Ignoring the pull from the global fit to $b \to s \mu \mu$ LNU can also stem from sizable BSM contributions to $b \to s e e$. In this case modes into final state electrons (and corresponding neutrinos) are dominant.

\mysection{Acknowledgements}
This project is supported in part 
by the {Bundesministerium f\"ur Bildung und Forschung (BMBF)}. We  thank Ivo 
de Medeiros Varzielas for useful discussions.

\end{document}